\documentclass[final,11pt,times,twocolumn]{elsarticle}
\usepackage[utf8]{inputenc}
\usepackage{amsmath,natbib}
\usepackage{mathrsfs}
\usepackage{graphicx}
\usepackage{epstopdf}
\usepackage{enumitem}
\usepackage{siunitx}
\usepackage{color}
\setlist{leftmargin=5cm}
\makeatletter
\def\ps@pprintTitle{%
   \let\@oddhead\@empty
   \let\@evenhead\@empty
   \def\@oddfoot{\reset@font\hfil\thepage\hfil}
   \let\@evenfoot\@oddfoot
}
\makeatother

\begin{document}

\begin{frontmatter}



\title{Capturing Strain Stiffening Using Volume Controlled Cavity Expansion}


\author[cee]{Shabnam Raayai-Ardakani}

\author[me,cee]{Tal Cohen\corref{cor1}}
\ead{talco@mit.edu}

\address[me]{Department of Mechanical Engineering, Massachusetts Institute of Technology, Cambridge, MA 02139}

\address[cee]{Department of Civil and Environmental Engineering, Massachusetts Institute of Technology, Cambridge, MA 02139}

\cortext[cor1]{Corresponding author.}

\begin{abstract}

\noindent Strain-stiffening is a well-documented behavior in soft biological materials such as liver and brain tissue. Measuring and characterizing this nonlinear response, which is commonly considered as a mechanism for damage prevention, is of great interest to engineers for design of better  biomimetic materials, and to physicians for diagnostic purposes. However, probing the elastic response of soft or biological materials at large deformation  in their natural habitat, is an arduous task. Here, we present the Volume Controlled Cavity Expansion (VCCE) technique as an \textit{in-vivo} measurement method that offers the ability of characterizing the stiffening response of materials in addition to identifying their shear modulus. By employing minimal constitutive representations involving only two constants (Mooney-Rivlin, Gent, and Ogden) we show that for the conventional PDMS samples, this technique and an accompanying data analysis method capture the shear modulus, as well as providing reliable measures of the stiffening behavior of the samples. 

\end{abstract}




\end{frontmatter}



\section{Introduction}
\label{sec:Introduction}

\noindent It has been well documented that strain-stiffening that may appear at large deformations in biological materials can effectively prevent damage \cite{yan2017duplicating}. For example, collagen-I networks \cite{motte2013strain}, lung tissue \cite{suki2011lung},  human aortic and mitral leaflets and chordae tendineae \cite{clark1973stress,grande2005mitral}, jellyfish \cite{gambini2012micro}, arteries \cite{shadwick1999mechanical}, cornea \cite{kampmeier2000thermal}, plasma clots \cite{shah1997strain}, fibrin gels \cite{shah1997strain, wen2007local}, brain \cite{mihai_family_2017, budday_mechanical_2015,Mihai_2015}, scar tissue and uninjured skin  \cite{Corr_2013,langer1978anatomy} are among the biological systems identified as strain-stiffening and are inspiring scientists to fabricate synthetic materials able to re-create similar responses \cite{yan2017duplicating}.  

In addition to biological materials, the interest in soft materials at large strains has been continuously increasing. Due to the introduction and expansion of several areas, such as soft-robotics \cite{kim2013soft,majidi2014soft, roche2014bioinspired}, flexible electronics \cite{harris2016flexible, liu2017lab, rogers2010materials}, Micro-Electro-Mechanical Systems (MEMS) \cite{ziaie2004hard}, , bio-materials \cite{ye2015supramolecular,patel2012review, sharma2014biomaterials}, controlling cell functions \cite{levental2007soft}, drug delivery \cite{lavan2003small, osada1993stimuli}, micro-fluidics \cite{ziaie2004hard}, adhesives \cite{de1996soft, chiche2005cavity, carelli2007effect}, and material instabilities  \cite{raayai2016mechanics, cerda2003geometry, li2012mechanics}, it is crucial to be able to characterize the response of soft materials to obtain more accurate predictions and designs. 

While conventional methods such as tensile testing and indentation are still vastly used for characterization of the soft and biological material, there is growing interest in minimally invasive \textit{in-vivo} techniques which do not require fabrication of specific sample shapes, to be tested outside of their original habitat. Cui et al. \cite{cui2011cavitation} have previously shown that \textit{ex-vivo} measurement of the mechanical properties of eye lens produces lower values compared to the \textit{in-vivo} results. In addition, minimally invasive \textit{in-vivo} measurements can assist with diagnosis of pathological conditions such as cancer \cite{samani2007inverse, paszek2005tensional}, atherosclerosis \cite{imura1988arteriosclerotic, hayashi1994aortic, wuyts1995elastic, claridge2009measurement,qiu2019biophysics}, sickle cell disease \cite{qiu2019biophysics}, fibrosis \cite{yeh2002elastic}, or glaucoma \cite{last2011elastic, russell2012elastic}, where the stiffening of the tissue is a bio-marker \cite{qiu2019biophysics}. Furthermore, conventional techniques are easily applicable to stiffer materials (shear modulus larger than 300 kPa), and thus characterization of softer materials using the conventional techniques can be challenging especially as we attempt to understand the mechanical response of the biological tissues \textit{in-vivo}.

Recently, the method of Volume Controlled Cavity Expansion (VCCE) was introduced in \cite{raayai2019volume} where the shear modulus is extracted through volume controlled injection of an incompressible fluid to expand a  cavity inside the material while measuring the internal pressure. The applicability of VCCE was confirmed  in soft PDMS (polydimethylsiloxane) samples, for which the neo-Hookean (NH) constitutive model is able to provide a reliable representation of the material behavior prior to failure. Later, in \cite{raayai2019Intimate}, it was shown that  using this volume controlled technique one can track the intertwined process of cavitaiton and fracture to obtain insight on the specific failure mechanism. The VCCE method was introduced to complement the Cavitation Rheology (CR) technique invented by Crosby and co-workers \cite{zimberlin2007cavitation,crosby2011blowing, zimberlin2010cavitation, zimberlin2010water, cui2011cavitation, hashemnejad2015nonlinear, hutchens2014soft, blumlein2017mechanical} by removing the need for \textit{a priori} assumptions on the constitutive response and the failure mechanism.   


When the effect of the finite chain extensibility is triggered, the response of soft and biological materials cannot be captured by the NH model. In this work, we aim to expand on the use of the VCCE for characterization of the fully nonlinear response  (prior to the onset of fracture). To do so, we focus on three constitutive models: Gent \cite{gent1996new}, Mooney-Rivlin (MR) \cite{mooney1940theory, rivlin1992elasticity,mihai2017characterize}, and one-term Ogden model \cite{ogden1972large}, and we write all the models as a function of two distinct parameters: the shear modulus $\mu$ and an additional dimensionless  parameter to identify the response in the larger stretches. 

Here, we perform the experiments and analysis for PDMS samples. While PDMS is widely used in soft materials and micro-fluidics research, there is no central data-base of the the hyper-elastic properties of PDMS at different mixing ratios (PDMS exhibits a very low loss tangent \cite{meng2011additive}). Therefore, in addition to simplicity of fabricating samples with various levels of stiffness by controlling the cross-linking density, we aim to also report the elastic properties of various PDMS samples based on their cross-linking ratios\footnote{This work is focused on Sylgard 184 samples, and PDMS samples from other companies \cite{larsen2003elastic} exhibit different responses.}.

In the next section we begin by comparing theoretical predictions of cavity expansion using the different constitutive models. The experimental method and data analysis technique are then described in section \ref{sec:exp}.  In section \ref{sec:results} we present the experimental results obtained using the different constitutive models to measure the shear modulus and an additional material parameter, and we discuss and compare the relative performance of the different models. Ultimately, in section \ref{sec:conclude} offer a few concluding remarks. 

\section{Theoretical Background}
\label{sec:theory}

\begin{figure*}[!ht]
    \centering
    \includegraphics[width = 1 \textwidth]{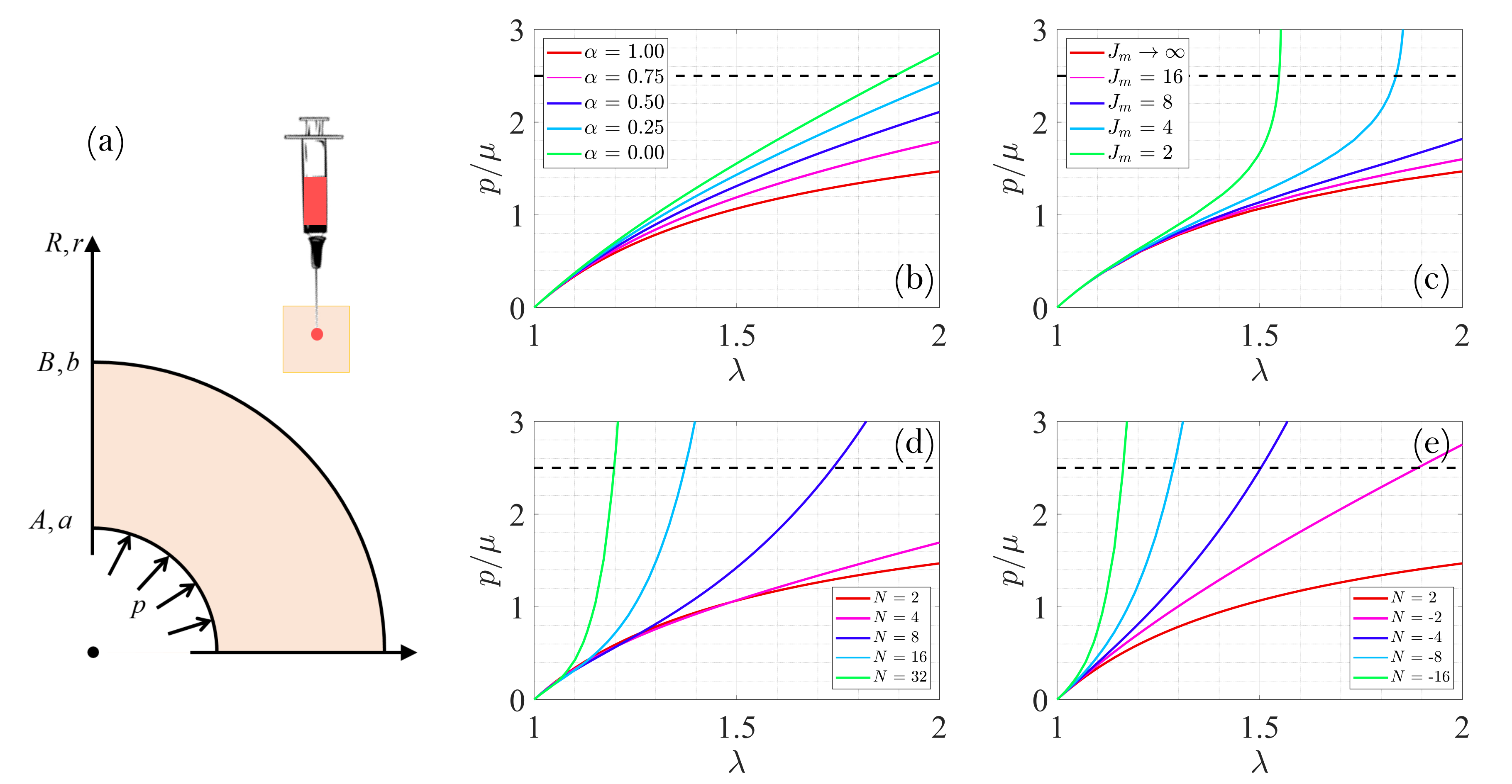}
    \caption{Theoretical response of hyperelastic materials in the cavity expansion problem derived using Eq.\@\ref{eq:cavity} with (a) Mooney Rivlin, (b) Gent, (c) Ogden with $N>0$, and (d) Ogden with $N<0$ constitutive relations. The cases of $\alpha = 1$, $J_m \rightarrow \infty$ and $N = 2$ captures the neo-Hookean model (shown in red). Dashed line at $p/\mu = 2.5$ corresponds to the NH cavitation limit.}
    \label{fig:cavity}
\end{figure*}

\noindent The problem of expanding a spherical cavity inside a body has been considered by various researchers for various constitutive models  \cite{gent1959internal, Gent_Tompkins_1969, lindsey1967triaxial, horgan1989cavity, chou1989cavitation, chou1989void, gent2005elastic, ganghoffer1995expansion, horgan1986bifurcation, chung1987note, horgan1992void, horgan1989void, horgan1995cavitation, polignone1993cavitation, polignone1993effects, hang1992cavitation, durban2004dynamic, cohen2009analysis, cohen2013fundamental, cohen2015dynamic}. In short, consider a spherical cavity of radii $A$ and $a$ in the undeformed and deformed configurations respectively, inside a body of incompressible material with outer radii $B$ and $b$ in the undeformed and deformed configurations, respectively (see Fig.\@\ref{fig:cavity}a). The strain energy density of the material can then be written in the form of $W(\lambda_{\theta})$ where $\lambda_{\theta} = r/R$ is the hoop stretch of the body. Therefore, the pressure inside the cavity can be found as \cite{raayai2019volume}
\begin{equation}
    {p} = \int\displaylimits_{\lambda_b}^{\lambda} \dfrac{W'(\lambda_{\theta})}{\lambda_{\theta}^3-1} d\lambda_{\theta}
    \label{eq:cavity}
\end{equation}
where $\lambda=a/A$ is the circumferential stretch at the cavity wall, $\lambda_b = b/B$ is the circumferential stretch at the outer radius of the body, and the prime denotes the first derivative with respect to $\lambda_{\theta}$. Using the constraint of incompressibility we can write the relation
\begin{equation}
    \lambda_b = \left( 1+ \left(\lambda^3-1 \right) \left(\dfrac{A}{B} \right)^3 \right)^{1/3}.
\end{equation}
If the radius of the cavity wall is much smaller than the radius of the body (i.e. $B/A \rightarrow \infty$) then $\lambda_b \to 1$. It was previously shown that within the range $1<\lambda<1.8$, the cavity expansion response of  NH bodies of $B/A>10$ are nearly identical \cite{raayai2019volume,raayai2019Intimate}.

The response of soft materials are described by various constitutive formulations, generically written based on the deformation gradient or simplified in terms of the principal stretches or the invariants of the Cauchy Green strain tensor. These models have been developed and enhanced to capture the nonlinear responses observed experimentally \cite{mihai2017characterize, boyce2000constitutive}. In the simplest form the widely used neo-Hookean (NH) model is defined based on the first invariant ($I_1$) of the left Cauchy-Green strain tensor \cite{mihai2017characterize} and a shear modulus, and can be shown to directly correspond with the Gaussian modeling of the entangled, cross-linked chains \cite{flory_principles_1953}. Considering an incompressible NH material with strain energy density of
\begin{equation}
    W(I_1) = \dfrac{\mu}{2} (I_1-3)
    \label{eq:NH}
\end{equation}
one can use Eq.\@\ref{eq:cavity} to re-derive the pressure-stretch formulation originally found by Gent and Lindley \cite{gent1959internal}
\begin{equation}
    \dfrac{p}{\mu} = \dfrac{5}{2} - \dfrac{2}{\lambda} - \dfrac{1}{2}\dfrac{1}{\lambda^4}.
    \label{eq:NH_cav}
\end{equation}
This result was  previously used  in \cite{raayai2019volume} for the analysis of VCCE data for soft PDMS samples ($\mu < 80$ [kPa]). However, the NH formulation can only capture the response of the material in terms of the shear modulus and applies for moderate strains. As one moves to larger deformations, the the finite length of the polymer chains limit this deformation (also known as the locking stretch) and alter the response of the material to deviate from NH behavior, toward a stiffening scenario. This strain-stiffening response can be modeled in various forms; while some have provided an extension to the NH form, in recent years many researchers have introduced more restricting statistical models \cite{horgan2006phenomenological,boyce2000constitutive} for the microscopic modeling of entangled chains such as the Wang-Guth 3-chain model \cite{wang1952statistical} and the Arruda-Boyce 8-chain model \cite{arruda1993three}. Here, we will focus on the cases of Mooney-Rivlin, Gent, and Ogden models. 

By addition of a second term, based on the second invariant ($I_2$) of the left Cauchy Green tensor, to the NH formulation, the Mooney-Rivlin (MR) model \cite{mooney1940theory, rivlin1992elasticity,mihai2017characterize} offers a broader range of responses and can capture a constant modulus in shear throughout the deformation process \cite{boyce2000constitutive}. The strain energy density of an incompressible MR material can be written as 
\begin{equation}
    W(I_1,I_2) = \dfrac{\mu}{2} \left[ \alpha (I_1-3) + (1-\alpha)(I_2-3) \right].
    \label{eq:MR}
\end{equation}
Here, $\alpha$ is a dimensionless parameter that distributes the action of the shear modulus  between the two terms in the MR formulation. If a material response is only dominated by the effect of the first invariant, then $\alpha = 1$ (i.e. NH), and if the response is controlled by the second invariant, then $\alpha = 0$ (see Fig.\@\ref{fig:cavity}b). Note that except for the case of $\alpha=1$ (i.e. NH), MR model does not predict an asymptotic response (cavitation limit) at large stretches. 

In a different form, Gent \cite{gent1996new}  proposed to extend the NH formulation and included an additional parameter $J_m$ to capture the maximum or limiting value of the first invariant of the Left Cauchy Green strain tensor as a result of the finite chain extensibility (i.e. locking stretch). The strain energy density of a Gent material can be written as 
\begin{equation}
    W(I_1) = \dfrac{\mu}{2}(I_1-3)J_m \ln \left( \dfrac{J_m}{J_m-I_1+3} \right)
    \label{eq:gent}
\end{equation}
and the pressure-stretch response of Gent materials with various values of $J_m$ are plotted using Eq.\@\ref{eq:cavity} in Fig.\@\ref{fig:cavity}c. It is clear from Eq.\@\ref{eq:gent}, that at the limit of $J_m \rightarrow \infty$, this model simplifies to the NH model and hence response (shown with the red curve) returns to the form of Eq.\@\ref{eq:NH_cav}. Through this expansion, the Gent model is able to capture the behavior of materials which are able to sustain large deformations without any visible stiffening $J_m \gtrsim 10$ as well as materials for which the chains lock at smaller stretches (as shown with $J_m = 2$ and $J_m=4$). The Gent response, except for the case of $J_m \rightarrow \infty$ (i.e. NH limit), does not have have a theoretical asymptote at $\lambda \rightarrow \infty$. This constitutive relation is able to capture the NH behavior at small strains as well as the strain stiffening at larger strains and it has been shown to be closely comparable with the Arruda-Boyce 8-chain model \cite{boyce1996direct}.

In a different expansion of NH and MR models, Ogden \cite{ogden1972large} has proposed a functional form in terms of summation of polynomials based on the principal stretches in the form of
\begin{equation}
    W(\lambda_1,\lambda_2,\lambda_3) = \sum_{i=1}^n \dfrac{2C_i}{N_i} (\lambda_1^{N_i} + \lambda_2^{N_i} + \lambda_3^{N_i} -3) ,
\end{equation}
where $C_i$ are the material coefficients, $N_i$ the exponent of the polynomials, and $\mu = \sum_{i=1}^n C_i N_i$ is the shear modulus. Here, we employ the simplest form of the Ogden model with only one term ($n=1$) and re-write the above equation as
\begin{equation}
     W(\lambda_1,\lambda_2,\lambda_3) = \dfrac{2\mu}{N^2} (\lambda_1^{N} + \lambda_2^{N} + \lambda_3^{N} -3),
     \label{eq:Ogden}
\end{equation}
which depends on the shear modulus and a dimensionless exponent $N$. Note that $N$ can take any real value and the case of $N=2$ returns the NH model while $N=-2$ returns the MR model with $\alpha = 0$ (see Fig.\@\ref{fig:cavity}d and \ref{fig:cavity}e)\footnote{The pressure-stretch responses are plotted separately for positive and negative $N$ for clarity.}. The Ogden model, with the wide range of available choices for the exponent $N$, has the ability to capture the response of a wide class of soft and biological materials. Ogden model has proven to be useful in characterizing the  behavior of soft biological materials, especially brain matter \cite{mihai2017family,mihai2017characterize, budday2017viscoelastic} and other organs \cite{umale2013experimental}. 

One key difference in the various models presented here is in the shape of the curves shown in Fig.\@\ref{fig:cavity}. If we call a curve with $\partial^2 p/\partial \lambda^2<0$ concave, one can see that the MR curves are all concave while Gent and Ogden models are able to capture responses with both concave and convex ($\partial^2 p/\partial \lambda^2>0$) forms. In addition, while MR offers more variations in the responses than the NH model ($\alpha = 1$), it still is limited when compared to the Gent and Ogden models.

A summary of the models and their respective constants are shown in Tbl.\@\ref{tbl:models}. Here, employing these constitutive models, we will analyze the pressure-volume VCCE measurements of various PDMS samples in terms of their shear modulus and stiffening behavior through the different coefficients listed in Tbl.\@\ref{tbl:models}.

\begin{table}[!ht]
\begin{center}
  \caption{Constitutive Models and Relevant Material Constants}
  \label{tbl:models}
  \begin{tabular}{llll}
    \hline
    Model  & Eq. & Constant(s)   \\
    \hline
    neo-Hookean (NH) &  \eqref{eq:NH} & $\mu$   \\
    Mooney-Rivlin (MR) & \eqref{eq:MR} & $\mu \ \& \ \alpha$   \\
    Gent &   \eqref{eq:gent} & $\mu \  \&  \ J_m$  \\
    Ogden & \eqref{eq:Ogden}& $\mu \ \& \ N$    \\
    \hline
  \end{tabular}
  \end{center}
\end{table}

\section{Experimental Methods and Data Analysis}
\label{sec:exp}
\noindent The VCCE experiments  in this work have been performed using the set-up and calibration developed by Raayai-Ardakani et al. \cite{raayai2019volume}. The experimental setup consists of two attachments to an Instron universal testing machine that transforms the machine into a volume controlled syringe pump. The experiments are performed by controlling the displacement of the syringe piston which linearly translates to controlling the volume of the dispensed fluid. Throughout each test, the force exerted on the piston is measured and then transformed to pressure using the calibration discussed earlier \cite{raayai2019volume}. All experiments were performed using four different volumetric rates of discharge (less than 5.5 $\mu \rm L/s$) and no differences have been observed. 

PDMS samples were fabricated using Sylgard 184 by mixing the two parts in various base:cross linking ratios. The samples were fabricated in cuboids of $\sim 30 \ \rm mm$ in length to ensure the samples are large enough that the outer size of the body does not interfere with the analysis of the results  (i.e. $\lambda_b \to 1$) \cite{raayai2019volume}. 

In our previous work \cite{raayai2019volume}, we have analyzed the pressure-volume measurements by employing a NH constitutive model to obtain the elastic modulus of soft PDMS samples (assuming incompressible $E \approx 3\mu$). In short, in a cavity expansion experiment, it is not possible to measure the initial volume (or radius) of the cavity made inside the body of material being tested. Therefore, by identifying the initial radius ($A$) as an unknown in the analysis, we introduced an algorithm to fit the experimental results to the constitutive relation and find the values of the shear modulus and the initial radius of the cavity. 

Here, we expand this methodology to obtain information regarding the material of interest, by including additional material constants in our data analysis steps. Therefore, we have extended our algorithm to fit any of the above introduced constitutive models to the collected data to find the shear modulus, the second constant ($\alpha$, $J_m$, or $N$), as well as the initial radius of the cavity. 

Our data fitting process \cite{raayai2019volume}, which is a search procedure based on trial and error, relies on the first derivative of the pressure with respect to $\lambda$ to correct the value of the guessed shear modulus used at every step. Using Eq.\@\ref{eq:cavity} with $\lambda_b = 1$ we can write
\begin{equation}
    \dfrac{\partial p}{\partial \lambda} = \dfrac{W'(\lambda)}{1-\lambda^3},
\end{equation}
and employing the chain rule we have
\begin{equation}
    \dfrac{\partial p}{\partial a} = \dfrac{1}{A}\dfrac{\partial p}{\partial \lambda}.
\end{equation}

Depending on the constitutive model, one can use the above result at a chosen point (with $a$ - calculated from the volume) to obtain a calculated guess for the value of the elastic modulus $\mu$ for each step in the trial and error process (for example the point of maximum slope was used in \cite{raayai2019volume}). Accordingly, the algorithm introduced previously \cite{raayai2019volume} is expanded to not only search for the shear modulus, but also to look for an additional parameter (dependent on the chosen constitutive model) where the combination of the two constants offer the lowest error between the measurements and the theoretical result within the range of interest. It should be noted that the fitting process has been applied to only the region of the data that exhibit an elastic response, prior to any indication of damage (drops in pressure). 

\section{Results and Discussion}
\label{sec:results}

\begin{figure*}[!ht]
    \centering
    \includegraphics[width  =0.94 \textwidth]{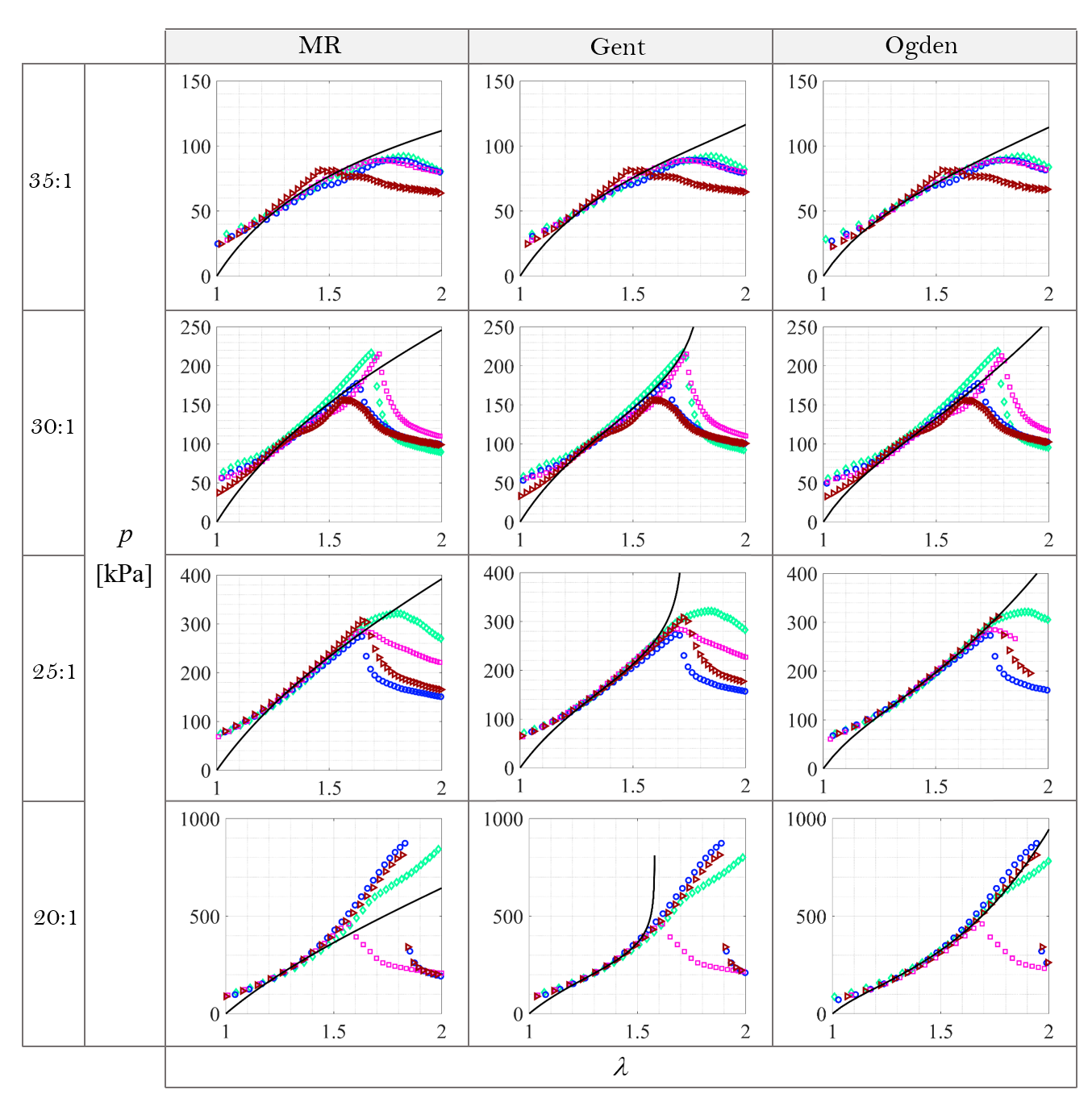}
    \caption{Response of various PDMS samples and their corresponding MR, Gent, and Ogden fits shown with black solid line. Four different colored symbols denote four separate tests. The corresponding fitted coefficients and their variations are listed in Tbl.\@\ref{tbl:PDMS_responses}. Note that the responses exhibit pre-stresses at $\lambda = 1$ which fade away after $\lambda \approx 1.1-1.2$. The fitting is focused on the range of data beyond $\lambda = 1.2$.}
    \label{fig:PDMS_responses}
\end{figure*}

\noindent Here, we measure the pressure-volume of the cavity during the expansion process in samples with various cross-linking ratios. Each set of data are analyzed separately using the three constitutive models of interest and a few of the representative results are presented in Fig.\@\ref{fig:PDMS_responses}. The plots in every row are the results for the same PDMS samples analyzed using the different models. To ensure repeatability, for each cross-linking density, four separate samples were tested and analyzed (different colored symbols) and in each fitting process a separate set of initial radius, shear modulus, and the second constant is calculated.\footnote{This results in slight difference between the figures in each row.} In each plot, one fitted line, calculated with the average of the shear modulus and the corresponding $J_m$, $\alpha$, or $N$, is shown with solid black line. The magnitude of the mean values as well as their variation (the $95\%$ confidence interval) are listed in Tbl.\@\ref{tbl:PDMS_responses} for reference.

Notice that the measured responses in Fig.\@\ref{fig:PDMS_responses} exhibit initial pre-stresses at $\lambda = 1$ which fade away beyond $\lambda \approx 1.1 - 1.2$. This pre-stress is related to the initial conditions created by needle insertion process \cite{rattan2019effect}, which is more pronounced for stiffer samples. By focusing on the entire range of data and considering theoretical curves for which  $p(\lambda=1)=0$, the fitting procedure is able to bypass this initial pre-stress and captures the nonlinear response at large strains. 


The discussion of the results are organized in the following format: first, the results of the analysis of the various samples using each of the constitutive models (following the columns of Fig.\@\ref{fig:PDMS_responses}) is presented. Afterwards, moving to the rows of Fig.\@\ref{fig:PDMS_responses}, the effectiveness of each of the constitutive models in capturing the key features of each of the samples is compared and contrasted. Then a summary of the measured elastic constants as a function of the cross-linking of the samples is presented and the trends are discussed. 

\begin{table*}[!ht]
\begin{center}
  \caption{Values of shear modulus, and other constants from fitting the measurements to various constitutive relations as shown in Fig.\@\ref{fig:PDMS_responses}.}
  \label{tbl:PDMS_responses}
  \begin{tabular}{|c |cc |cc |cc|}
    \hline
     & \multicolumn{2}{c|}{MR } & \multicolumn{2}{c|}{Gent} & \multicolumn{2}{c|}{Ogden} \\
    \cline{2-7}
     Sample  & $\mu$ [kPa] & $\alpha$ & $\mu$ [kPa] & $J_m$ & $\mu$ [kPa] & $N$ \\
    \hline
       $35:1$ & $67.29 \pm 2.04$ & $0.85 \pm 0.15$ & $67.12 \pm 1.13$ & $9.50 \pm 4.80$ & $66.43 \pm 1.14$ & $4.47 \pm 1.17$ \\
      $30:1$ & $114.88 \pm 2.46$ & $0.48 \pm 0.13$ & $116.04 \pm 2.34$ & $3.70 \pm 0.24$ & $115.12 \pm 1.48$ & $5.63 \pm 0.54$ \\
       $25:1$ & $163.61 \pm 2.14$ & $0.28 \pm 0.10$ & $161.55 \pm 2.51$ & $3.05 \pm 0.19$ & $163.19 \pm 2.87$ & $6.28 \pm 0.21$ \\
       $20:1$ & $234.17 \pm 2.83$ & $0.00 \pm 0.00$ & $233.41 \pm 0.68$ & $2.15 \pm 0.10$ & $234.68 \pm 2.10$ & $7.73 \pm 0.22$ \\
    \hline
  \end{tabular}
  \end{center}
\end{table*}


First, focusing on the MR model (the left column of Fig.\@\ref{fig:PDMS_responses}), one can only find acceptable fittings for 35:1, 30:1 and 25:1 samples and for stiffer sample of 20:1, MR can only offer a moderate fit up to $\lambda<1.45$ and capturing $\alpha=0$ which is the ultimate level of stiffening possible in MR. This results in underestimating the pressure response at larger stretches as well as inability to capture the convex form of the measurements, especially for stiffer PDMS samples (20:1) (Note in the gradual change in the curvature of the measurements from concave to convex as the cross-linking density of the samples is increased).


Unlike MR, Gent is able to capture both the concave and the convex form of the responses (the middle column of Fig.\@\ref{fig:PDMS_responses}). However, for samples of about $\mu>150$ [kPa] (25:1 and 20:1) one can see that the limit inherent in $J_m$ as defined in the model, constrains it from fitting to the entire elastic portion of the measurements and is only applicable up to a maximum stretch dictated by the magnitude of $J_m$. As a result, for such samples, we can obtain a reasonable fit using the Gent model, but only up to $\lambda \approx 1.5-1.6$ and beyond that the model drifts apart from the measurements. 


In contrast, the one-term Ogden model, is able to offer fits for a wider range of measurements. As seen in the right hand column in Fig \ref{fig:PDMS_responses}, the Ogden model captures the entire elastic response of the samples (prior to the point where fracture is inevitable). By tuning the exponent of the principal stretches, the Ogden model, even with one term is able to represent the responses very well, especially for 20:1 samples were both MR and Gent can offer limited representations.


Comparing the models, only for the softer 35:1 sample (first row of Fig.\@\ref{fig:PDMS_responses}), the three models capture nearly identical functional forms, very close to the NH response. However, for stiffer samples (other rows), the three models capture different fits, where only Gent and Ogden can capture the convex form of the results and only Ogden fitting the entire range of elastic measurements. 


Unlike the rest, 35:1 samples present a larger scatter in the measured $J_m$. This is due to the fact that the four different samples capture a wide range of responses with $J_m$ from 4 to 14. Thus, for this sample, the cross-linking density is not low enough to present a fully NH behavior and it is not high enough to be fully stiffening 
(like the rest of the samples). As a result, we observe responses on both sides of the spectrum (see Tbl.\@\ref{tbl:PDMS_responses}). However, as seen in Fig.\@\ref{fig:cavity}, the difference between the curves with $4<J_m<14$ within $1.4<\lambda<1.6$ are very small compared to the magnitude of the shear modulus of the samples and can be at times indistinguishable as seen in Fig.\@\ref{fig:PDMS_responses}. 

\begin{figure}[!ht]
    \centering
    \includegraphics[width = 0.45 \textwidth]{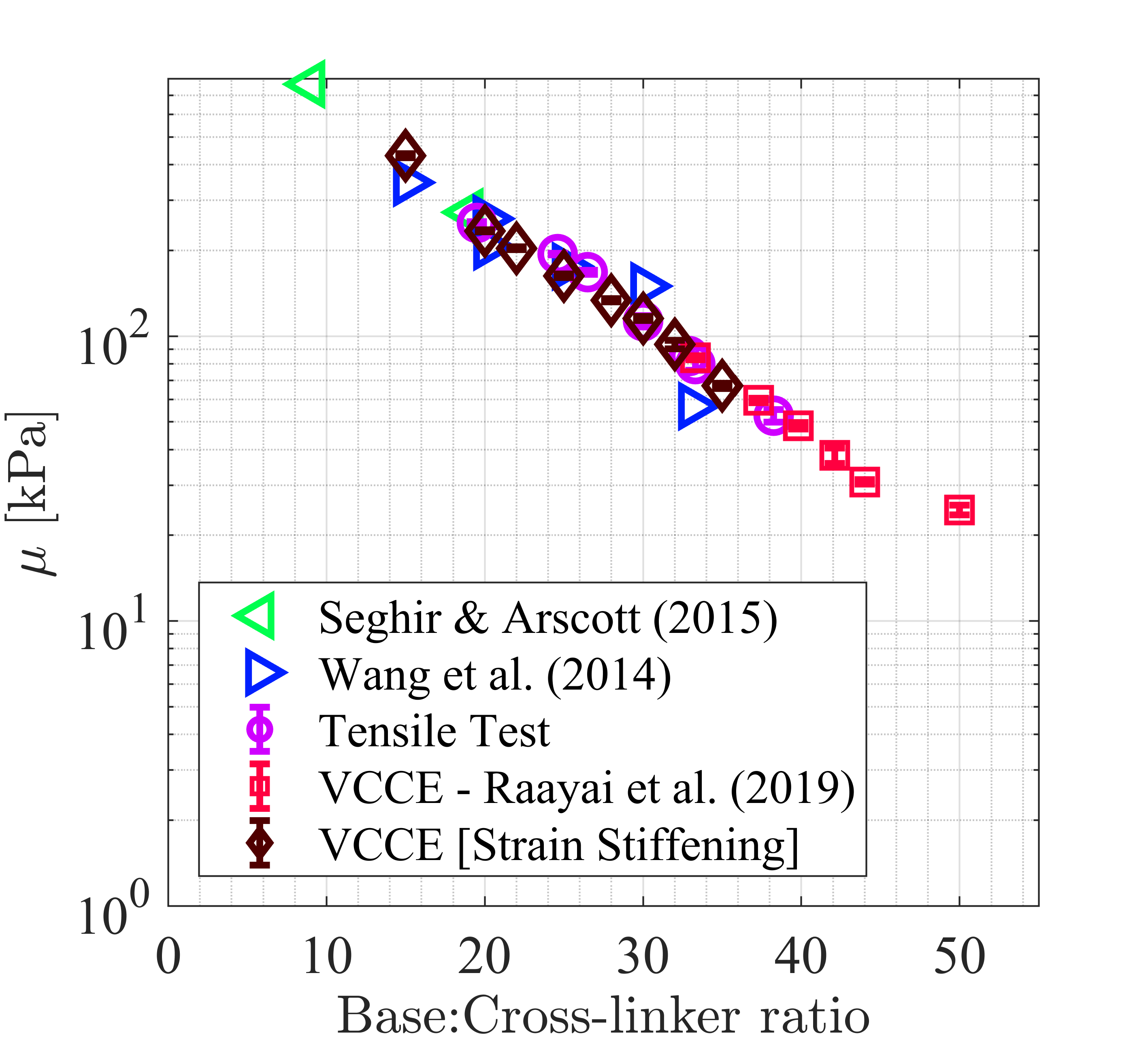}
    \caption{Variation of the average shear modulus measured for PDMS samples as a function of the cross-linking density, in comparison with our previous measurements \cite{raayai2019volume}, as well as the measurements of Wang et al. \cite{wang2014crosslinking} and Seghir and Arscott \cite{seghir2015extended}.}
    \label{fig:mu}
\end{figure}
\begin{figure}[!ht]
    \centering
    \includegraphics[width = 0.45 \textwidth]{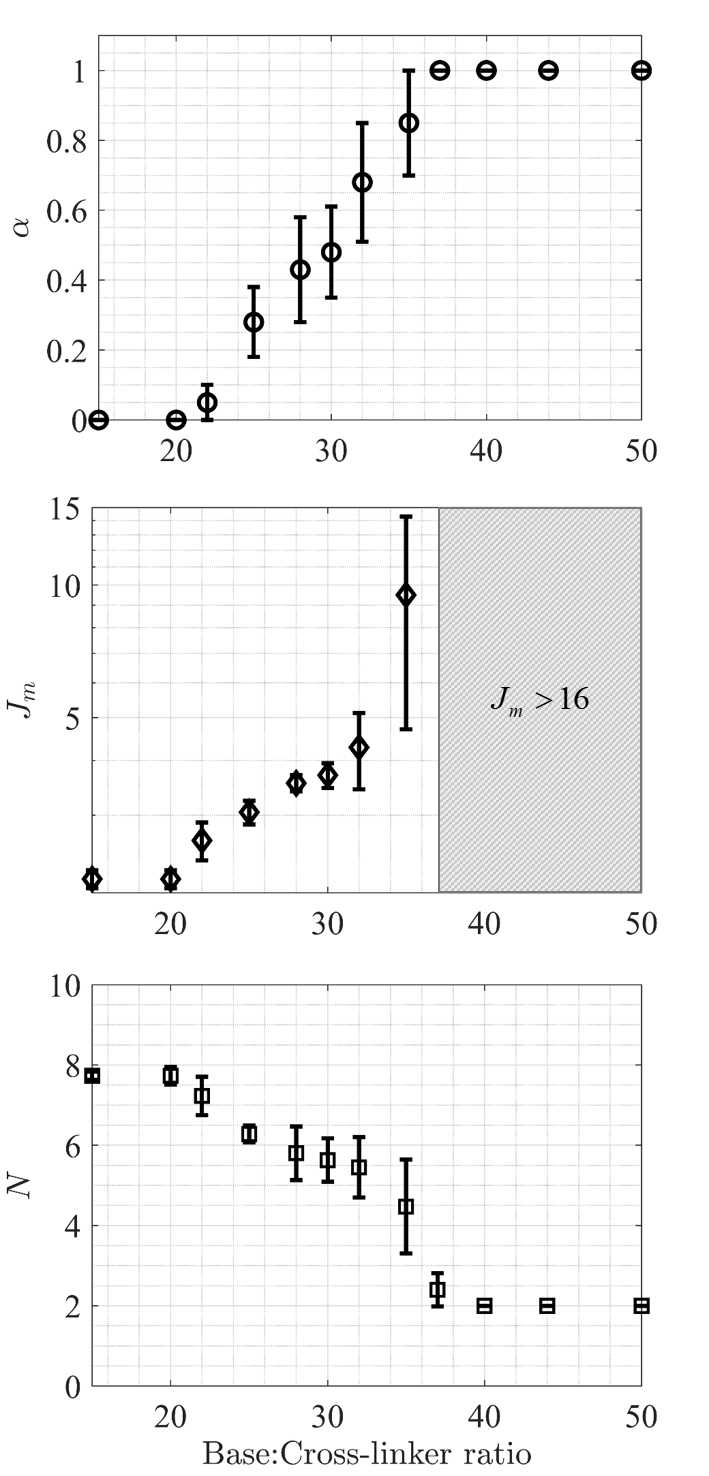}
    \caption{Variation of the material constants $\alpha$, $J_m$ (semi-logarithmic), and $N$ measured for PDMS samples as a function of the cross-linking density.}
    \label{fig:alpha_J_N}
\end{figure}

A similar analysis has been repeated for other PDMS samples within the range of 50:1 and 15:1 and a summary of the measured constants (shear modulus, $\alpha$, $J_m$, and $N$ for all the samples are plotted in Figs. \ref{fig:mu} and \ref{fig:alpha_J_N} against the cross-linking density of the samples. As it can be seen (and expected), increasing the cross-linking density of the samples results in a consistent increase in the shear modulus and the results of our VCCE measurements agree with tensile tests \cite{raayai2019volume} as well as the data reported by other researchers \cite{wang2014crosslinking, seghir2015extended}.


On the softer side (40:1 and lower), Ogden and MR capture $N=2$ and $\alpha = 1$, and if analyzed with a Gent model, one finds $J_m>20$, all confirming a fully NH behavior (for $1<\lambda<1.7$). PDMS 37:1 samples reveal $J_m$ values larger than 16 where as seen in Fig.\@\ref{fig:cavity}c, Gent model with such large values of $J_m$ exhibits a response nearly identical to the NH for $1<\lambda<1.6$. Similarly MR can only show $\alpha = 1$, i.e. NH behavior and when analyzed against the Ogden model, it is possible to capture an exponent of $N = 2.40 \pm 0.14$, again nearly identical to the NH response.\footnote{The plots of the measurements of these three samples can be found in \cite{raayai2019volume}.}


Focusing on Fig.\@\ref{fig:alpha_J_N}, for the rest of the samples in the range of 35:1-20:1, as the cross-linking density is increased, we consistently observe an enhancement in the stiffening response captured through the variations in the second constants. All three models allow us to capture a gradual change in the second coefficient, however, as it can be seen in Fig.\@\ref{fig:alpha_J_N}a, MR captures larger scatters as large as $20\%$ (compared to entire range of $\alpha$ available). With Gent, we capture a gradual decrease in $J_m$ toward $J_m \approx 2$ for samples with cross-linking density of 32:1 or higher. However, the Gent model is unable to capture the pressure-stretch responses beyond $\lambda \approx 1.5-1.6$ for samples of 25:1 and higher densities. Similarly, with the Ogden model, we can capture a gradual increase in $N$ from $N=2$ toward a value of $N\approx 8$ for samples of 20:1 and higher capturing the entire range of measured elastic data. 

Unfortunately, there are no reports of measurements of such second constants reported in literature for comparable samples. Seghir and Arscott \cite{seghir2015extended} use a generalized MR to analyze their results, but only report the shear modulus as presented in Fig.\@\ref{fig:mu}. Kim et al. \cite{kim2011measurement} report softer responses for PDMS films of 15:1  (less than half the shear modulus measured here) and thus a fair comparison is not possible. Larsen et al. \cite{larsen2003elastic} report the MR coefficients for their PDMS samples, but since the samples were fabricated not using Sylgard 184, the results are not directly comparable. In addition, due to the local-triaxial nature of this test, comparison of these measured coefficients with those calculated from conventional uniaxial and biaxial methods can prove to be difficult and future work should focus on combining protocols for testing of soft materials in the nonlinear regime in different stress states. In particular, among the constitutive models used here, the most appropriate model for a given material should be chosen by conducting tests in  additional stress states \cite{mihai2017characterize}.


Lastly, we note here that PDMS has been employed in this study due its common usage in both engineering applications and soft matter research. Additionally, it allows for easy adjustment of stiffness through the change in the composition, thus making it possible to investigate the range of capability of the VCCE technique in measuring the nonlinear response of the materials.

\section{Conclusion}
\label{sec:conclude}
\noindent We have presented the VCCE technique as a method for measuring not only the shear modulus, but also characterizing the strain stiffening response of the soft materials. Through this volume-controlled method, we have measured the volume-pressure response obtained by expansion of a cavity at the tip of a needle inside the material of interest. Then, by fitting the results to three different constitutive models, we measure two separate elastic constants representing the shear modulus and the stiffening response. 

By focusing on models involving only two constants, we expand the algorithm introduced previously to search for both these constants and the initial radius of the cavity, through a search for curves offering the lowest residual between the measurements and models. We have implemented this for Mooney-Rivlin, Gent, and one-term Ogden models applied to measurements in PDMS samples of various base:cross-linking ratios. It is shown that increasing the cross-linking density in the system results in a continuous increase in both the shear modulus of the samples and the strain-stiffening response, with the results exhibiting a convex form. Using all three models we are able to see a gradual enhancement in the stiffening response through changes in the corresponding model parameters (i.e. $\alpha$, $J_m$ and $N$). 

Among the three different models, we notice that Ogden has the ability to capture the response of the results over nearly the entire elastic range of the measurements, while Gent per its definition, is limited depending on the value of $J_m$. As a result, while we were able to capture $J_m$ values offering good agreements with the experiments, they were limited to a range of $\lambda<1.5-1.6$. Similarly, MR was able to only offer concave-form fits to the measurements and for stiffer samples (20:1 and higher), only offering a limited fit up to $\lambda \approx 1.45$ with $\alpha = 0$ which is its utmost stiffening option.

Finally, it is shown that the VCCE technique is able to provide repeatable  measurement of nonlinear pressure-stretch response of materials, in accordance with common constitutive models. In future this technique can be applied to measure the response of more complex soft and biological materials \textit{in-vivo}, and can aid in quantification of material properties for bio-mechanics research and medical diagnosis.

\section*{Acknowledgement}
\noindent TC wishes to acknowledge the support of the Army Research Office and Dr. Ralph A. Anthenien, Program Manager, under award number W911NF-19-1-0275.

\section*{References}
\bibliographystyle{elsarticle-num-names}
\biboptions{compress}
\bibliography{library}

\end{document}